Report on the EAS 2021 Symposium

# Exploring the High-Redshift Universe with ALMA

held online, 28–29 June 2021

Evanthia Hatziminaoglou[1]
Gergö Popping[1]
Martin Zwaan[1]

[1] ESO

The properties of the interstellar medium (ISM) of the highest-redshift galaxies and quasars provide important indications of the complex interplay between the accretion of baryons onto galaxies, the physics that drives the build-up of stars out of this gas, the subsequent chemical evolution and feedback processes and the reionisation of the Universe. The Atacama Large Millimeter/submillimeter Array (ALMA) continues to play a pivotal role in the characterisation of the ISM of high-redshift galaxies. Observations of the dust continuum emission, atomic fine-structure and molecular lines arising from high-redshift galaxies are now carried out routinely, providing ever more constraints on the theoretical models of galaxy formation and evolution in the early Universe. The European Astronomical Society's EAS 2021 symposium dedicated to the exploration of the high-redshift Universe with ALMA provided a forum for the observational and theoretical high-redshift ALMA communities to exchange their views and recent results in this rapidly evolving field.

The EAS 2021 symposium talks addressed five intertwined themes spread over six sessions: ALMA's unmatched potential to study high-redshift galaxies, its contribution to our understanding of the properties of dust in the early Universe, the properties of the ISM, the feeding and feedback processes at high redshift and the synergies with current and future facilities. Figure 1 shows a schematic representation of some examples of science with ALMA. Highlights from each theme are summarised below, underlining the power of ALMA and its remarkable contribution to groundbreaking discoveries.

## High-redshift galaxies as observed by ALMA

Observations of "old" galaxies at $z > 8$, an epoch at which the peak of dust emission falls in ALMA's bands 7 and 8, can provide information about the time at which these very first galaxies were born. Indeed, ALMA observations of galaxies at $z \sim 9$ confirmed a formation redshift beyond $\sim 11$ (Hashimoto et al., 2018; Laporte et al., 2021a), providing an indirect probe into the cosmic dawn.

According to Mahsa Kohandel, simulations predict that early galaxies could form their discs as early as during the epoch of reionisation (EoR) but Sander Schouws showed that the majority of galaxies (~ 70%) at the EoR are dispersion-dominated, with stellar mass only accounting for up to 10% of their dynamical mass. Ana Carolina Posses Nascimento, on the other hand, presented a galaxy at $z = 6.8$ that is rotationally supported, as derived from [CII] observations. At $z \sim 4$ at least some galaxies have regular rotating discs (as explained by Francesca Rizzo); however, Fernanda Roman de Oliveira showed that there is a diversity in $V/\sigma$, leading to the well established dominance of rotation by $z \sim 2$.

Other than targeted observations of high-$z$ galaxies, several ALMA extragalactic Large Programs[1] have demonstrated ALMA's capability to be used as a survey instrument, despite its small field of view. Indeed, many of the results presented in the symposium come from such projects. Hanae Inami focused her invited talk on the Reionization Era Bright Emission Lines Survey (REBELS; Bouwens et al., submitted to ApJ), a [CII] and [OIII] survey of 40 Lyman-break galaxies (LBGs) at $6.5 < z < 9.5$ and reported dust continuum detections for about half of the sources. The [CII] and dust emissions are spatially extended and diverse (Inami, in preparation; also reported by Rebecca Bowler in her talk). Serendipitous detections of dust-obscured galaxies, discussed by Pascal Oesch, imply that normal star-forming galaxies existed in the EoR but were essentially missed until now, setting constraints on the obscured star formation out to the EoR.

In a talk introducing the ALMA Large Programme to INvestigatE (ALPINE; Le Fèvre et al., 2020), which carried out [CII] observations of 122 main-sequence star-forming galaxies at $z \sim 5$, Matthieu Bethermin reported on the large morpho-kinematic diversity amongst $z \sim 5$ main-sequence galaxies. ALPINE demonstrated the presence of large gas reservoirs at $z \sim 6$ and a marginal decrease with redshift of the gas consumption timescale from $z \sim 6$ to the present day (Dessauges-Zavadsky et al., 2020). The same study showed that the progenitors of Milky Way-like galaxies and massive present-day galaxies evolve differently across cosmic time. Finally, ALPINE suggests that the star formation rate density remains almost constant from $z \sim 2$ to $z \sim 6$ and is five to eight times higher than the optical/UV estimates (Gruppioni et al., 2020; Figure 2).

Figure 1. ALMA's windows on the Epoch of Reionization (EoR) — from Joris Witstok's talk.

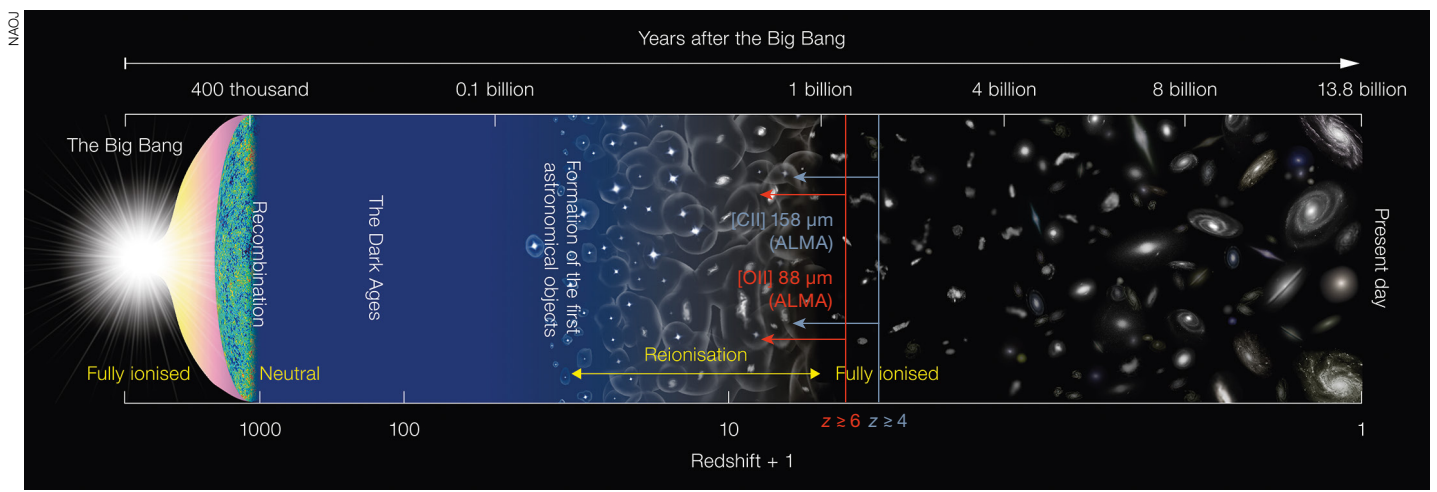

ALMA has resolved more than half of the sources contributing to the cosmic infrared background; however, the study of the sub-milliJansky population is still difficult, even with ALMA's capabilities. Kotaro Kohno presented results from the ALMA Cluster Survey (Caputi et al., 2021; Fujimoto et al., 2021; Laporte et al., 2021b), a Band 6 survey of 33 clusters from the Hubble Space Telescope (HST) Treasury programme that explores the faint-end regime of the 1-millimetre ALMA sky and that has yielded 133 unique > 5σ continuum detections at 1 millimetre and about 60 line detections. The power of lensing in the study of high-$z$ galaxies with ALMA was also demonstrated by Francesca Rizzo, who reported on the presence of bulges already at these redshifts (Rizzo et al., 2021).

### Dust in the early Universe

Dust plays a crucial role in galaxy assembly despite contributing only 1% of the total baryonic mass. ALMA is now called upon to constrain dust temperature and dust production mechanisms at redshifts beyond four. The Calzetti law (Calzetti, Kinney & Storchi-Bergmann, 1994) has traditionally been used to correct for dust attenuation in galaxies across cosmic time. ALMA, however, showed a rapid evolution of the properties of dust at higher redshifts, where galaxies agree more with a Small Magellanic Cloud attenuation curve, representative of lower-metallicity environments.

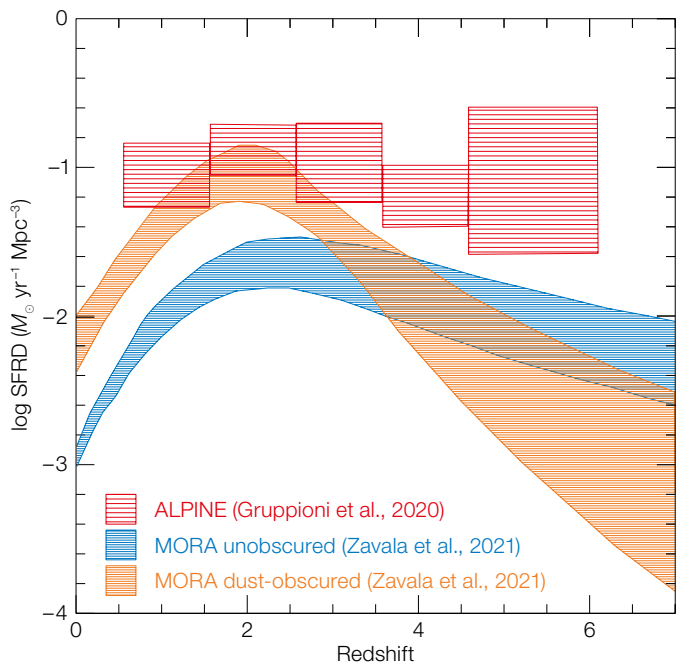

Figure 2. Star formation history of the Universe: obscured, derived from infrared/submillimetre observations (orange) and unobscured, derived from UV surveys (blue) and the ALPINE Large Programme (red).

Although theoretical work supports hotter dust (> 60 K and as high as 90 K) at higher redshifts, ALMA multi-band observations of four $z \sim 6$ LBGs lead to temperature estimates of $T = 30–43$ K (Faisst et al., 2020), despite degeneracies in the dust emissivity index (the exponent of the power-law fit) and large uncertainties. This is in very good agreement with the latest REBELS results of 39–52 K (Sommovigo et al., in preparation). Joris Witstok also reported the flattening of the $T$–$z$ relation up to $z \sim 7$.

At the highest redshifts ($z \sim 8$) the ALMA detections of very dust-rich galaxies (for example, Laporte et al., 2017; Tamura et al., 2019) challenge the models of dust production. Such large dust masses could be the result of high supernova rates or efficient grain growth in the ISM, more efficient in massive, more metal-rich galaxies, as reported by Elisabete da Cunha (see also the review by Hodge & da Cunha, 2020). Additionally, a more top-heavy IMF, as suggested by Zhi-Yu Zhang based on ALMA studies of CO isotopologues in $z \sim 3$ galaxies, could also lead to the efficient production of dust. All the above indicate the clear need for multi-band, high-frequency ALMA observations to measure temperature, luminosity and dust mass in galaxies (da Cunha et al., 2021) and quasars (outlined in Roberta Tripodi's talk) at the EoR.

### The ISM in the early Universe

ALMA observations of rest-frame far-infrared fine-structure lines such as [CII] 158 µm and [OIII] 88 µm can provide direct measurements of the molecular gas content, metallicity, star formation rate (SFR) and gas density of high-$z$ galaxies. Figure 3 is a schematic representation of the various ISM components from which the fine structure and molecular line emission originate.

[CII] 158 µm, the brightest line in most star-forming galaxies, is used to trace star formation across cosmic time. $L_{[CII]}$–SFR$_{UV+IR}$ at $z > 6$ is consistent with that in the local Universe, but the dispersion is larger (Carniani et al., 2020). Yuexing Li reported on the theoretical luminosity functions derived by combining Adaptive Refinement Tree (ART2) and the TNG100 and 300 cosmological simulations, that reproduce the $L_{[CII]}$–SFR$_{UV+IR}$ relation all the way to $z \sim 10$, suggesting that [CII] is indeed a good tracer of star formation but pointing out that the correlation is not linear at these high redshifts.

Joint [CII]-[OIII] line detections are powerful diagnostics of the conditions in the ISM at the EoR, with $L$[OIII]/$L$[CII] larger than in the local Universe (Carniani et al., 2020). The deviation from the local relation might be due to the more extended [CII] emission with respect to that of [OIII]/UV (Ginolfi et al., 2020a; Herrera-Camus et al., 2021). With the help of zoom-in cosmological simulations and ISM modelling, the above can be put in a wider context, indicating ISM conditions pointing towards an efficient conversion of gas into stars. The high metallicity and gas density are indicative of a dense ISM where bursts of star formation rapidly enrich the gas (Vallini et al., 2021). Simulations by Ramos Padilla et al. (2021), combining EAGLE cosmological simulations with CLOUDY cooling tables, showed that at $z \sim 6$ [CII], [OIII] and other lines in fact trace different phases of the ISM depending on metallicity and SFR.

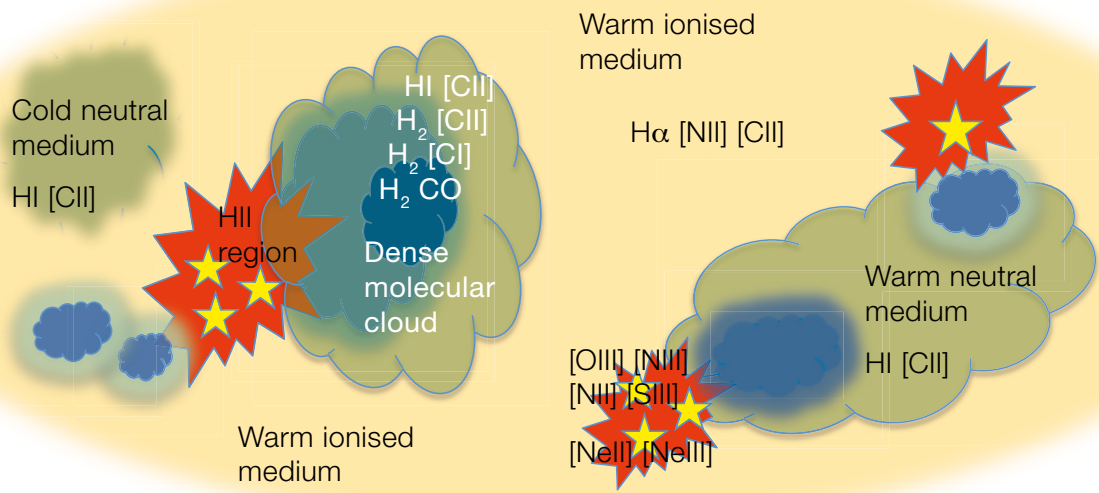

Figure 3. Origin of fine structure and molecular line emission from the various ISM components. Reproduced from van der Tak et al. (2018) with the author's permission.

Extended [CII] emission observed in galaxies beyond redshift $\sim 6$ (although not seen in a deep stack by Jean-Baptiste Jolly) can be produced by cooling outflows with mass loading factors larger than three (for example, Pizzati et al., 2020). The presence of outflows is supported by larger values of $L$[OIII]/$L$[CII] compared to those in the local Universe (large values also reported by Joris Witstok) seen in galaxies at $z > 6$. However, the loading factors are observed to be below or around one, indicating that the gas in the outflow might be in a warm ionised phase, or that the surface brightness of the outflows is low (Carniani et al., in preparation).

### Feeding and feedback at high redshift

The circumgalactic medium (CGM) is a place where physical processes such as mixing, cooling and heating happen, modifying the properties of the gas that is moving in or out of the galaxies. One of the main challenges currently is to explain the coexistence of the cold gas ($T < 10^4$ K) in a medium filled with hot gas. "*The lack of cold gas in CGM simulations is a bias rather than a prediction*," to quote Claudia Cicone. Higher-resolution simulations lead to larger fractions of cold gas, but simulations are still not converging.

Owing to the lack of sensitivity of the current facilities, molecular gas is not detected in the CGM at $z \sim 0$. ALMA, however, can observe the CGM at high redshift. Claudia Cicone reported on observations of a halo extending out to a radius of 200 kpc around a $z = 2.2$ X-ray-selected quasar (Cicone et al., 2021). There is no evidence of an overdensity around it, making the interpretation of this CO halo challenging.

Overdensities around $z \sim 6$ quasars turn out to be low-$z$ ($\sim 1$–2) objects (Meyer, in preparation), in agreement with [CII] incidence rates of companion overdensities reported in the literature.

The recent literature on the stacking of [CII] observations in high-$z$ active galactic nuclei (AGN) in search of evidence for outflows reports contradictory results. Even though the different spatial resolutions might explain the tension if outflows extend out to several kiloparsecs, Roberto Maiolino postulated that the estimated high-$z$ cold outflows are not powerful enough to quench star formation via ejection, indicating either the importance of ionised outflows or a weaker coupling with the ISM.

At the same time, ALPINE found signatures of star formation-driven outflows (i.e., broad wings) in the stacked profiles of [CII] in galaxies with SFRs > 25 $M_\odot$ yr$^{-1}$ (the median SFR of the ALPINE sample), but no evidence for outflows amongst galaxies with lower SFR (Ginolfi et al., 2020a). The outflowing gas in the highest-SFR galaxies moves at a maximum velocity ($v_{out} \sim 500$ km s$^{-1}$) below the escape velocities (400–800 km s$^{-1}$) and it will be trapped in the CGM of these galaxies (Ginolfi et al., 2020b). Hydrogen fluoride was reported by Maximilien Franco to be another excellent probe of outflowing molecular gas, as well as a good tracer of $H_2$ in galaxies. Finally, ALMA observations of a $z \sim 5.5$ main-sequence galaxy showed ISM properties similar to those of local starburst galaxies, with extended [CII] emission, evidence for an outflow and a regular rotating disc (Herrera-Camus et al., 2021).

### Synergies with current and future facilities

ALMA has undoubtedly revolutionised our knowledge of the physics of the early Universe. Synergies with existing and future facilities help push ALMA's potential to its limits. From the discovery of the very first galaxies to galaxy kinematics at redshifts close to the EoR to metallicities at redshifts of six and beyond, the possibilities for further groundbreaking discoveries are almost endless.

One of ALMA's challenges is to resolve the gas properties of galaxies with well resolved dynamics at high redshift and to decompose galaxies into their constituents (for example, Lelli et al., 2021). An obvious synergy, brought out by Mark Swinbank in his invited talk, is to improve dynamical constraints of galaxies at high redshifts by combining their ALMA CO and [CII] maps with stellar mass maps measured by the James Webb Space Telescope (JWST) and kiloparsec-scale dynamics measured by instruments like the Enhanced Resolution Imager and Spectrograph (ERIS) on ESO's Very Large Telescope and, in the future, the High Angular Resolution Monolithic Optical and Near-infrared Integral field spectrograph (HARMONI) on ESO's Extremely Large Telescope.

The JWST is going to observe optical rest-frame spectra of $z > 6$ galaxies for the first time. Combined observations of strong optical lines ([OII], [OIII], [SII] doublet) with the JWST and [OIII] 88 μm with ALMA will give direct estimates of metallicities at $z > 6$ (Jones et al., 2020).

ALMA [OIII] 88 μm observations impose strong constraints on the spectral energy distribution fit of Spitzer/IRAC-selected $z \sim 7$–9 galaxy candidates (Roberts-Borsani, Ellis & Laporte, 2020), especially prior to the arrival of the JWST, breaking the degeneracies between nebular emission and Balmer break. Further ALMA-Spitzer synergies are illustrated by the Spitzer Matching survey of the Ultra-VISTA deep Stripes (SMUVS) project (using the Infrared Array Camera channels 1 & 2 to observe the Ultra-VISTA deep stripes) presented by Tomoko Suzuki, that studies the submillimetre properties of high-redshift galaxies in the A³COSMOS catalogue (Liu et al., 2019). Galaxies with ALMA counterparts are found to be systematically massive ($M_\star > 10^{10}\,M_\odot$) and tend to have larger dust extinction and higher star formation activity. Additionally, ALMA, Multi Unit Spectroscopic Explorer (MUSE) and HST observations of luminous galaxies at $z > 6$ presented by Jorryt Matthee indicate that high-redshift luminous galaxies likely reside in early ionised bubbles (Matthee et al., 2020).

Meanwhile, it remains unclear how the supermassive black holes (with masses often above $10^9\,M_\odot$) residing in the centres of more than 200 known quasars at $z > 6$ have reached these sizes. Current theories predict a massive gas reservoir and ALMA-supported studies of these objects point towards rapidly star-forming high molecular gas and dust content hosts. These are often accompanied by submillimetre companions but simulations presented by Fabio Di Mascia indicate that such systems might be, instead, quasar merging systems. Alyssa Drake reported on MUSE Ly-α halo observations and complementing ALMA cold gas and gas continuum observations to find decoupled kinematics between halo gas and the ISM in high-$z$ quasars (Drake et al., 2020). Follow-up observations of ALMA-detected high-redshift galaxies with the JWST or the Origins Space Telescope might reveal dust-obscured or faint AGN residing in them.

### Conclusions

In its first decade of operation, ALMA has revolutionised our understanding of the high-redshift Universe. It has demonstrated the presence of massive dusty systems that contain large gas reservoirs fuelling vigorous star formation and accretion processes. ALMA has also exposed the presence of cold and metal-enriched gas in the haloes surrounding galaxies at $z > 4$, indicative of CGM pollution by outflows. ALMA has furthermore opened entire new windows onto the chemistry and ISM physics of galaxies during the EoR, contributing to key questions pertaining to the build-up and reionisation of the early Universe. This symposium was a brilliant demonstration of these groundbreaking results and served as an inspiration to continue to push the envelope to deliver more exciting ALMA high-redshift results in the decade to come.

### Acknowledgements

In addition to the authors of this article, the Scientific Organising Committee of the symposium included Paola Andreani, Andy Biggs, Gabriela Calistro Rivera, Jackie Hodge, Rob Ivison, Kirsten Kraiberg Knudsen, Renske Smit, Remco van der Burg and Eelco van Kampen. Thanks to them, and to all the speakers and participants, the symposium was a great success.

### References

Calzetti, D., Kinney, A. L. & Storchi-Bergmann, T. 1994, ApJ, 429, 582
Caputi, K. I. et al. 2021, ApJ, 908, 146
Carniani, S. et al. 2020, MNRAS, 499, 5136
Cicone, C. et al. 2021, A&A, 654, L8
da Cunha, E. et al. 2021, ApJ, 919, 30
Dessauges-Zavadsky, M. et al. 2020, A&A, 643, A5
Drake, A. B. et al. 2020, ApJ, 902, 37
Faisst, A. L. et al. 2020, MNRAS, 498, 4192
Fujimoto, S. et al. 2021, ApJ, 911, 99
Ginolfi, M. et al. 2020a, A&A, 633, A90
Ginolfi, M. et al. 2020b, A&A, 643, A7
Gruppioni, C. et al. 2020, A&A, 643, A8
Hashimoto, T. et al. 2018, Nature, 557, 392
Herrera-Camus, R. et al. 2021, A&A, 649, A31
Hodge, J. A. & da Cunha, E. 2020, RSOS, 7, 200556
Jones, T. et al. 2020, ApJ, 903, 150
Laporte, N. et al. 2017, ApJL, 837, L21
Laporte, N. et al. 2021a, MNRAS, 505, 3336
Laporte, N. et al. 2021b, MNRAS, 505, 4838
Le Fèvre, O. et al. 2020, A&A, 643, A1
Lelli, F. et al. 2021, Science, 371, 713
Liu, D. et al. 2019, ApJ, 887, 235
Matthee, J. et al. 2020, MNRAS, 492, 1778
Pizzati, E. et al. 2020, MNRAS, 495, 160
Ramos Padilla, A. F. et al. 2021, A&A, 645, A133
Rizzo, F. et al. 2021, MNRAS, in press
Roberts-Borsani, G. W., Ellis, R. S. & Laporte, N. 2020, MNRAS, 497, 3440
Tamura, Y. et al. 2019, ApJ, 874, 27
Vallini, L. et al. 2021, MNRAS, 505, 5543
van der Tak, F. F. S. et al. 2018, PASA, 35, 2
Zavala, J. A. 2021, ApJ, 909, 165

### Links

[1] ALMA Large Programmes: https://almascience.org/alma-data/lp